\documentclass[12pt]{iopart}

\newcommand{\PLA}{{\it Phys. Lett.} A }
\newcommand{\PRA}{{\it Phys. Rev.} A }

\newcommand{\OC}{{\it Opt. Commun.} }

\newcommand{\UQ}{ARC Centre of Excellence for Quantum-Atom Optics, 
School of Physical Sciences, University of Queensland, Brisbane, 
Qld 4072, Australia.}
\usepackage{graphicx}% eps files
\usepackage{bm}% bold math
\usepackage{iopams}
\usepackage{setstack}
\usepackage[dvips]{color}
\begin{document}

\title[Bright entanglement]{Bright continuous-variable entanglement from the quantum optical dimer}

\author{M~J Mallon, M~D Reid and M~K Olsen}
\address{\UQ}

\begin{abstract}

We show theoretically that two evanescently coupled $\chi^{(2)}$ second harmonic generators
inside a Fabry-Perot cavity provide a tunable source of quadrature
squeezed light, Einstein-Podolsky-Rosen correlations and quantum entanglement.
Unlike systems using coupled downconverters, second harmonic generation has no oscillation threshhold, so that the entangled fields become
macroscopically occupied as soon as the pumping fields are turned on. This system also gives two frequencies at which the entangled fields can have macroscopic intensity.
We show how the entanglement properties can be controlled by adjusting the pumping, the coupling strengths
and the cavity detunings.
 
\end{abstract}


\submitto{\JPB}

\ead{mko@physics.uq.edu.au}

\maketitle
%-----------------------------------------------------------------------------------------------------------

\section{Introduction}

The system of two evanescently coupled $\chi^{(2)}$ processes operating in the second harmonic generation (SHG) regime inside a pumped Fabry-Perot cavity has been theoretically analysed by Bache \etal~\cite{dimer}, in terms of suppression of number difference fluctuations between the output modes. In this work the name quantum optical dimer was given to this device. Although the calculations performed by Bache \etal were sufficient to show that strong intensity correlations should exist between the outputs, they did not calculate any of the phase-dependent correlations which are necessary to demonstrate the existence of continuous variable bipartite quantum entanglement. In later works, Olsen and Drummond~\cite{mkopdd} and Olivier and Olsen~\cite{Nicolas}
have shown theoretically that a related device, operating in the downconversion regime of optical parametric oscillation, can provide a robust source of continuous variable entanglement and be used to demonstrate the 
Einstein-Podolsky-Rosen (EPR) paradox in both the above and below threshold regimes. In this work we will examine the quantum optical dimer proposal of Bache \etal in terms of the production of entanglement and EPR states. In principle this sytem may have all the operational advantages of the coupled downconverter proposals, along with the 
property that all output modes are always macroscopically excited in second harmonic generation since there is no oscillation threshold. 

Generally, the device we are proposing may be considered as either a 
single nonlinear crystal pumped 
by two spatially separated lasers, or two waveguides with a $\chi^{(2)}$ component. We
calculate phase-dependent correlations between the
outputs of the cavity, evaluating entanglement criteria due to Duan \etal~\cite{Duan}, Simon~\cite{Simon}, the lograithmic negativity~\cite{Goofrey}, as well as the Reid-Drummond EPR correlations~\cite{eprMDR,rd1,rd2} which can also be used to prove entanglement. Generically, our system is related to the nonlinear coupler, which is the name given to a system of two coupled waveguides
without an optical cavity by Pe\u{r}ina \etal~\cite{coupler}.
The device generally consists of two parallel optical waveguides
which are coupled by an evanescent overlap of the guided modes. The
quantum statistical properties of this device when the nonlinearity
is of the $\chi^{(3)}$ type have been theoretically investigated,
predicting energy transfer between the waveguides~\cite{Ibrahim} and the generation of correlated squeezing~\cite{korea}. When operated inside an optical cavity, entanglement between the output modes has been predicted~\cite{nlc2006}.
Here we show that the system with $\chi^{(2)}$ nonlinearity, and operating in the upconversion regime, is also potentially
an easily tunable source of single-mode squeezing and entangled states for both the low frequency fundamental modes and the high frequency up-converted harmonic modes. The spatial separation of
the output modes means that they do not have to be separated by optical
devices before measurements can be made, along with the unavoidable
losses which would result from this procedure. We will show that the correlations are tunable by controlling some of the
operational degrees of freedom of the system, including the evanescent
couplings between the two waveguides, the cavity damping rates, the input powers and the cavity
detunings. 

\section{Entanglement measures}
\label{sec:emaranhamento}

Before we begin to analyse the actual physical system, we will define and outline the measures which we will use below to demonstrate continuous-variable bipartite entanglement.
Entanglement is a property of quantum mechanics which is related to the inseparability of the combined density matrix of a system into density matrices for its subsystems. In the present situation, we will be interested interested in continuous variable bipartite entanglement between the output modes from each side of the dimer, which we shall label as $1$ and $2$.
It is firstly necessary to give the definition of the optical quadratures we will use, as the exact form of any inequalities depends on this. For the description of this system, we require four intracavity bosonic annihilation operators, $\hat{a}_{1},\,\hat{a}_{2},\,\hat{b}_{1},\,\hat{b}_{2}$, where the $\hat{a}_{j}$ are for the low frequency (fundamental) modes and the $\hat{b}_{j}$ are for the high frequency (harmonic) modes. We define the quadrature operators at the phase angle $\theta$ as 
\begin{equation}
\hat{X}_{a,j}^{\theta}=\hat{a}_{j}\e^{-i\theta}+\hat{a}_{j}^{\dag}\e^{i\theta},
\label{eq:quaddef}
\end{equation}
so that $[\hat{X}_{a,j}^{\theta},\hat{X}_{a,k}^{\theta+\pi/2}]=2i\delta_{jk}$, where $\delta_{jk}$ is the Kronecker delta, and similarly for the high frequency quadratures. The Heisenberg uncertainty principle (HUP) then requires $V(\hat{X}_{a,j}^{\theta})V(\hat{X}_{a,k}^{\theta+\pi/2})\geq 1\delta_{jk}$. In the interests of notational simplicity, we will label the quadrature $\hat{X}_{a,k}^{\theta+\pi/2}$ as $\hat{Y}_{a,k}^{\theta}$.

The EPR paradox stems from a famous
paper published in $1935$~\cite{EPR}, which used a gedanken experiment with particles which were entangled in position and momentum to show that local realism
was not consistent with the completeness of quantum mechanical theory. A direct and
feasible demonstration of the EPR paradox with continuous variables
was first suggested using nondegenerate parametric amplification~\cite{eprquad1,eprquad2,eprquad3}. This was possible because the optical quadrature phase amplitudes
used in these proposals have the same mathematical properties as the position and momentum
originally used by EPR. Even though the correlations between these
are not perfect, they are still entangled sufficiently to allow for
an inferred violation of the uncertainty principle, which is equivalent
to the EPR paradox~\cite{eprMDR,rd1,rd2}. An experimental demonstration
of this proposal by Ou \etal soon followed, showing a clear
agreement with quantum theory~\cite{Ou}. A recent theoretical proposal has examined a demonstration with the atomic field quadratures of massive particles, using the coherent dissociation of molecular Bose-Einstein condensates~\cite{rejection}.

Although the concept of entanglement is required to formulate the EPR paradox, for completeness we will outline the details of a proof given by Reid~\cite{PDbook} that seeming violations of a Heisenberg uncertainty principal (HUP) as in the experiment of Ou \etal~\cite{Ou} are automatically demonstrations of entanglement. We begin by assuming that a given system is bipartite separable and divide it into two subsystems $A$ and $B$. We now consider observables $\hat{x}^{A}$ and $\hat{y}^{A}$ of subsystem $A$, obeying the Heisenberg uncertainty principal with $V(\hat{x}^{A})V(\hat{y}^{A})\geq 1$. We now introduce $V_{inf}(\hat{x}_{A})$ as the measured error in the prediction for the outcome of a measurement $\hat{x}^{A}$ at $A$, based on a result at $B$, and similarly for $V_{inf}(\hat{y}_{A})$. The first task is to show that separability always demands $V_{inf}(\hat{x}^{A})V_{inf}(\hat{y}^{A})\geq 1$, so that violation of this inequality requires inseparability and hence entanglement of $A$ and $B$. The conditional probability of result $x^{A}$ for a measurement of $\hat{x}^{A}$ at $A$ given a simultaneous measurement of $\hat{x}^{B}$ at $B$ with result $x_{i}^{B}$ is
\begin{equation}
P(x^{A}|x_{i}^{B}) = \frac{P(x^{A},x_{i}^{B})}{P(x_{i}^{B})},
\label{eq:mdrEPR1}
\end{equation}
where, assuming separability
\begin{equation}
P(x^{A},x_{i}^{B}) = \sum_{r}P(r)P(x_{i}^{B}|r)P(x^{A}|r),
\label{eq:mdrEPR2}
\end{equation}
with a separable density matrix being written as
\begin{equation}
\rho = \sum_{r}P(r)\rho_{r}^{A}\rho_{r}^{B}.
\label{eq:mdrEPR3}
\end{equation}
In this case, $|x^{A}\rangle$ and $|x^{B}\rangle$ are eigenstates of $\hat{x}^{A}$ and $\hat{x}^{B}$, with $P(x^{A}|r)=\langle x^{A}|\rho_{r}^{A}|x^{A}\rangle$ and $P(x^{B}|r)=\langle x^{B}|\rho_{r}^{B}|x^{B}\rangle$. The mean of this conditional distribution is
\begin{eqnarray}
\eqalign{
\mu_{i} &= \sum_{x^{A}}x^{A}P(x^{A}|x_{i}^{B})\\
&= \frac{\sum_{r}P(r)P(x_{i}^{B}|r)\langle x^{A}\rangle_{r}}{P(x_{i}^{B})},
}
\label{eq:mdrEPR4}
\end{eqnarray}
where $\langle x^{A}\rangle_{r}=\sum_{x^{A}}P(x^{A}|r)$. The variance, $V_{i}(x)$, of the distribution $P(x^{A}|x_{i}^{B})$ is then
\begin{equation}
V_{i}(x) = \frac{\sum_{r}P(r)P(x_{i}^{B}|r)\sum_{x^{A}}P(x^{A}|r)(x^{A}-\mu_{i})^{2}}{P(x_{i}^{B})}.
\label{eq:mdrEPR5}
\end{equation}
For each $r$, the mean square deviation, $\sum_{x^{A}}P(x^{A}|r)(x^{A}-d)^{2}$, is minimised by the choice $d=\langle x^{A}\rangle_{r}$, so that for the choice $d=\mu_{i}$,
\begin{eqnarray}
\eqalign{
V_{i}(x) &\geq \frac{\sum_{r}P(r)P(x_{i}^{B}|r)\sum_{x^{A}}P(x^{A}|r)(x^{A}-\langle x^{A}\rangle_{r})^{2}}{P(x_{i}^{B})} \\
&= \frac{\sum_{r}P(r)P(x_{i}^{B}|r)V_{r}(x^{A})}{P(x_{i}^{B})},
}
\label{eq:mdrEPR6}
\end{eqnarray}
where $V_{r}(x^{A})$ is the variance of $P(x^{A}|r)$. We may also define a measured error, $V_{inf,est}(\hat{x}^{A})$, resulting from linear inference, which will not be better than that based on knowledge of the conditional probabilities, so that
\begin{eqnarray}
\eqalign{
V_{inf,est}(\hat{x}^{A}) &\geq V_{inf}(\hat{x}^{A}) \\
& \geq \sum_{x_{i}^{B}}P(x_{i}^{B})\frac{\sum_{r}P(r)P(x_{i}^{B}|r)V_{r}(x^{A})}{P(x_{i}^{B})} \\
& = \sum_{r}P(r)V_{r}(x^{A})\sum_{x_{i}^{B}}P(x_{i}^{B}|r) \\
& = \sum_{r}P(r)V_{r}(x^{A}). 
}
\label{eq:mdrEPR7}
\end{eqnarray}
We also note that $V_{inf}(\hat{y}^{A})\geq \sum_{r}P(r)V_{r}(y^{A})$, where $P(y^{A}|r)=\langle y^{A}|\rho_{r}^{A}|y^{A}\rangle$ and $|y^{A}\rangle$ is the eigenstate of $\hat{y}^{A}$. The Cauchy-Schwartz inequality then implies that 
\begin{eqnarray}
\eqalign{
V_{inf}(\hat{x}^{A})V_{inf}(\hat{y}^{A}) &\geq \sum_{r}P(r)V_{r}(x^{A})\times \sum_{r}P(r)V_{r}(y^{A}) \\
&\geq |\sum_{r}P(r)V_{r}(x^{A})V_{r}(y^{A})|^{2}.
}
\label{eq:mdrEPR8}
\end{eqnarray}
For any physical $\rho_{r}$, the HUP requires that $V_{r}(x^{A})V_{r}(y^{A})\geq 1$. For a separable state it is therefore required that
\begin{equation}
V_{inf}(\hat{x}^{A})V_{inf}(\hat{y}^{A})\geq 1.
\label{eq:mdrEPR9}
\end{equation}
As the measured errors are always at least as large as the inferred errors, this means that an observation of 
$V_{inf,est}(\hat{x}^{A})V_{inf,est}(\hat{y}^{A})<1$ is sufficient to prove inseparability and hence bipartite entanglement. We note that this violation is a sufficient but not a necessary condition, so that continuous variable bipartite entanglement may be present which is not detected by this means. An example of such a situation, which can occur for mixed states, is given by Bowen \etal~\cite{Warwick}. We also note here that the EPR paradox signifies a stronger form of entanglement, as has been discussed recently by Wiseman \etal in the context of steering~\cite{Wiseman}.

In practice any EPR measurement is usually done by defining orthogonal quadratures via a minimisation of the errors in a linear inference process~\cite{ndturco}. The apparent violation of the HUP for these quadratures then signals entanglement between the modes of the system. 
We note here that it has also been shown by Tan~\cite{Sze} that the existence of two orthogonal quadratures, the product of whose variances violates the limits set by the HUP, provides evidence of entanglement. Tan demonstrated this in the context of teleportation, with the outputs from a nondegenerate OPA mixed on a beamsplitter. This procedure has been extended to the case of tripartite entanglement~\cite{tripart} and its extension to larger numbers of modes is straightforward, even though the number of inequalities to be violated increases due to the different classes of inseparability which then exist~\cite{casos}.

The second of the entanglement measures is due to Duan \etal~\cite{Duan} and also Simon~\cite{Simon}, who developed inseparability criteria which are necessary and sufficient for Gaussian states, and sufficient in general. These criteria have recently been shown to be special cases of an infinite series of inequalities based on the non-negativity of determinants of matrices constructed from certain combinations of operator moments~\cite{Vogel}. In the general case, we may define combined quadrature operators similarly to Duan as
\begin{eqnarray}
\hat{X}_{\pm}^{\theta} &=& |b|\hat{X}_{1}^{\theta}\pm\frac{1}{|b|}\hat{X}_{2}^{\theta},\nonumber\\
\hat{Y}_{\pm}^{\theta} &=& |b|\hat{Y}_{1}^{\theta}\pm\frac{1}{|b|}\hat{Y}_{2}^{\theta},
\label{eq:XYplusminus}
\end{eqnarray}
where $b$ is an arbitrary non-zero real number. It may be shown that, for separable states,
\begin{equation}
V(\hat{X}_{\pm}^{\theta})+V(\hat{Y}_{\mp}^{\theta})\geq 2\left(b^{2}+\frac{1}{b^{2}}\right),
\label{eq:critDuan}
\end{equation}
with any violation of this inequality therefore demonstrating the presence of bipartite entanglement. In what follows, we will choose $b=1$ so that the lower bound of the inequality is $4$. While this is not the optimal choice for the general case, it is the appropriate choice for this system, due to the symmetry between modes $1$ and $2$.

The third measure which we will apply is the logarithmic negativity, proposed by Vidal and Werner as a computable measure of entanglement, as opposed to others which can be difficult to calculate~\cite{Goofrey}. We note here that this measure is defined for Gaussian states, to which we are also limited here due to the linearisation process which we will be using. We first define the system covariance matrix as
\begin{equation}
{\cal C}=\left[\begin{array}{cc}
{\cal C}_{1} & {\cal C}_{12} \\
{\cal C}_{21} & {\cal C}_{2}
\end{array}\right],
\label{eq:covmat}
\end{equation}
where
\begin{equation}
{\cal C}_{j}=\left[\begin{array}{cc}
V(\hat{X}_{j}^{\theta}) & V(\hat{X}_{j}^{\theta},\hat{Y}_{j}^{\theta}) \\
V(\hat{Y}_{j}^{\theta},\hat{X}_{j}^{\theta}) & V(\hat{Y}_{j}^{\theta})
\end{array}\right],
\label{eq:jcovmat}
\end{equation}
and
\begin{equation}
{\cal C}_{ij}=\left[\begin{array}{cc}
V(\hat{X}_{i}^{\theta},\hat{X}_{j}^{\theta}) & V(\hat{X}_{i}^{\theta},\hat{Y}_{j}^{\theta}) \\
V(\hat{Y}_{i}^{\theta},\hat{X}_{j}^{\theta}) & V(\hat{Y}_{i}^{\theta},\hat{Y}_{j}^{\theta})
\end{array}\right].
\label{eq:ijcovmat}
\end{equation}
Defining
\begin{equation}
\xi = \sqrt{(\det{\cal C}_{1}-\det{\cal C}_{12})-\sqrt{(\det{\cal C}_{2}-\det{\cal C}_{12})^{2}-\det{\cal C}}}
\end{equation}
the logarithmic negativity is then defined as
\begin{displaymath}
{\cal F}(\xi) = \left\{\begin{array}{ll}
-\log_{2}\xi & \mbox{if}\:\: \xi<1 \\
0 & \mbox{otherwise.}
\end{array}\right.
\end{displaymath}
Any non-zero value of ${\cal F}(\xi)$ is then an indication that the two modes are entangled. An interesting feature of this measure is that it has no dependence on quadrature angle, but becomes a function of frequency only. This shows that the logarithmic negativity is useful for demonstrating that continuous variable entanglement exists in a given system, but does not tell us at which quadrature angles the system may exhibit the necessary properties for uses such as teleportation.

\section{The system and equations of motion}
\label{sec:equations}

The physical device we wish to examine is the same as that described
in reference \cite{dimer}. As this device has been described there 
in detail, we will give a briefer description
of the essential features here. The system consists of two
coupled nonlinear $\chi^{(2)}$ waveguides inside a driven optical
cavity, which may utilise integrated Bragg reflection for compactness.
Each waveguide supports two resonant or near resonant modes at frequencies $\omega_{a}$ (fundamental) and $\omega_{b}$ (harmonic),
where $2\omega_{a}\simeq\omega_{b}$. The lower frequency modes at
$\omega_{a}$ are driven coherently with external laser fields, while
the nonlinear interaction within the waveguides produces second harmonic photons with frequency $\omega_{b}$. We assume that
only the cavity modes at these two frequencies are important and that
there is perfect phase matching inside the media. The two waveguides
are evanescently coupled. We will be interested in the phase-dependent
correlations necessary for an unambiguous demonstration of entanglement and the EPR paradox,
rather than intensity correlations considered in reference \cite{dimer}.

The effective Hamiltonian for the system can be written as 
\begin{equation}
{\mathcal{H}}_{eff}={\mathcal{H}}_{int}+{\mathcal{H}}_{couple}+{\mathcal{H}}_{pump}+{\mathcal{H}}_{res},
\label{eq:Heff}
\end{equation}
where the nonlinear interactions with the $\chi^{(2)}$ media are described
by 
\begin{equation}
{\mathcal{H}}_{int}=\rmi\hbar\frac{\kappa}{2}\left[\hat{a}_{1}^{\dag\;2}\hat{b}_{1}-\hat{a}_{1}^{2}\hat{b}_{1}^{\dag}+
\hat{a}_{2}^{\dag\;2}\hat{b}_{2}-\hat{a}_{2}^{2}\hat{b}_{2}^{\dag}\right]\,\,.
\label{eq:Hnl}
\end{equation}
In the above $\kappa$ denotes the effective nonlinearity of the waveguides
(we assume that the two are equal), and $\hat{a}_{k},\;\hat{b}_{k}$
are the bosonic annihilation operators for quanta at the frequencies
$\omega_{a},\;\omega_{b}$ within the nonlinear medium $k\;(=1,2)$. The coupling
by evanescent waves is described by 
\begin{equation}
{\mathcal{H}}_{couple}=\hbar J_{a}\left[\hat{a}_{1}\hat{a}_{2}^{\dag}+\hat{a}_{1}^{\dag}\hat{a}_{2}\right]+\hbar J_{b}
\left[\hat{b}_{1}\hat{b}_{2}^{\dag}+\hat{b}_{1}^{\dag}\hat{b}_{2}\right],
\label{eq:Hcouple}
\end{equation}
where the $J_{k}$ are the coupling parameters at the two frequencies,
as described in reference \cite{dimer}, where it
is stated that the lower frequency coupling, $J_{a}$, is generally
stronger than the higher frequency coupling, $J_{b}$, and also that
values of $J_{a}$ as high as $50$ times the lower frequency cavity
loss rate may be physically reasonable. The cavity
pumping is described by 
\begin{equation}
{\mathcal{H}}_{pump}=\rmi\hbar\left[\epsilon_{1}\hat{a}_{1}^{\dag}-\epsilon_{1}^{\ast}\hat{a}_{1}+\epsilon_{2}\hat{a}_{2}^{\dag}-
\epsilon_{2}^{\ast}\hat{a}_{2}\right],
\label{eq:Hpump}
\end{equation}
where the $\epsilon_{k}$ represent pump fields which we will describe
classically. Finally, the cavity damping is described by 
\begin{equation}
{\mathcal{H}}_{res}=\hbar\sum_{k=1}^{2}\left(\Gamma_{a}^{k}\hat{a}_{k}^{\dag}+\Gamma_{b}^{k}\hat{b}_{k}^{\dag}\right)+h.c.,
\label{eq:Hres}
\end{equation}
where the $\Gamma^{k}$ represent bath operators at the two frequencies
and we have made the usual zero temperature and Markov approximations for the
reservoirs.

With the standard methods~\cite{GardinerQN}, and using the operator/c-number
correspondences $(\hat{a}_{j}\leftrightarrow\alpha_{j},\hat{b}_{j}\leftrightarrow\beta_{j})$,
the Hamiltonian can be mapped onto a Fokker-Planck equation for the
Glauber-Sudarshan P-distribution~\cite{Roy,Sud}. However, as the
diffusion matrix of this Fokker-Planck equation is not positive-definite,
it cannot be mapped onto a set of stochastic differential equations.
Hence we will use the positive-P representation~\cite{plusP} which,
by doubling the dimensionality of the phase-space, allows a Fokker-Planck
equation with a positive-definite diffusion matrix to be found and
thus a mapping onto stochastic differential equations. Making the
correspondence between the set of operators $(\hat{a}_{j},\hat{a}_{j}^{\dag},\hat{b}_{j},\hat{b}_{j}^{\dag})$
$(j=1,2)$ and the set of c-number variables $(\alpha_{j},\alpha_{j}^{+},\beta_{j},\beta_{j}^{+})$,
we find the following set of equations, 
\begin{eqnarray}
\eqalign{
\frac{\rmd\alpha_{1}}{\rmd t} = \epsilon_{1}-(\gamma_{a}+\rmi\Delta_{a})\alpha_{1}+\kappa\alpha_{1}^{+}\beta_{1}+\rmi J_{a}\alpha_{2}+\sqrt{\kappa\beta_{1}}\;\eta_{1}(t), \\
\frac{\rmd\alpha_{1}^{+}}{\rmd t} = \epsilon_{1}^{\ast} -(\gamma_{a}-\rmi\Delta_{a})\alpha_{1}^{+}+\kappa\alpha_{1}\beta_{1}^{+}-\rmi J_{a}\alpha_{2}^{+}+
\sqrt{\kappa\beta_{1}^{+}}\;\eta_{2}(t), \\
\frac{\rmd\alpha_{2}}{\rmd t} = \epsilon_{2}-(\gamma_{a}+\rmi\Delta_{a})\alpha_{2}+\kappa\alpha_{2}^{+}\beta_{2}+\rmi J_{a}\alpha_{1}+\sqrt{\kappa\beta_{2}}\;\eta_{3}(t), \\
\frac{\rmd\alpha_{2}^{+}}{\rmd t} = \epsilon_{2}^{\ast}-(\gamma_{a}-\rmi \Delta_{a})\alpha_{2}^{+}+\kappa\alpha_{2}\beta_{2}^{+}-\rmi J_{a}\alpha_{1}^{+}+
\sqrt{\kappa\beta_{2}^{+}}\;\eta_{4}(t), \\
\frac{d\beta_{1}}{\rmd t} = -(\gamma_{b}+\rmi\Delta_{b})\beta_{1}-\frac{\kappa}{2}\alpha_{1}^{2}+\rmi J_{b}\beta_{2}, \\
\frac{d\beta_{1}^{+}}{\rmd t} = -(\gamma_{b}-\rmi\Delta_{b})\beta_{1}^{+}-\frac{\kappa}{2}\alpha_{1}^{+\;2}-\rmi J_{b}\beta_{2}^{+}, \\
\frac{d\beta_{2}}{\rmd t} = -(\gamma_{b}+\rmi\Delta_{b})\beta_{2}-\frac{\kappa}{2}\alpha_{2}^{2}+\rmi J_{b}\beta_{1}, \\
\frac{d\beta_{2}^{+}}{\rmd t} = -(\gamma_{b}-\rmi\Delta_{b})\beta_{2}^{+}-\frac{\kappa}{2}\alpha_{2}^{+\;2}-\rmi J_{b}\beta_{1}^{+},}
\label{eq:PPSDE}
\end{eqnarray}
where the $\gamma_{k}$ represent the cavity damping rates at each frequency. We have also added
cavity detunings $\Delta_{a,b}$ from the two resonances, so that
for a pump laser at frequency $\omega_{L}$, we have $\Delta_{a}=\omega_{a}-\omega_{L}$ and
$\Delta_{b}=\omega_{b}-2\omega_{L}$. Below, in section~\ref{sec:detune},
we will investigate detuning effects in greater detail. The real Gaussian
noise terms have the correlations $\overline{\eta_{j}(t)}=0$ and
$\overline{\eta_{j}(t)\eta_{k}(t')}=\delta_{jk}\delta(t-t')$. Note
that, due to the independence of the noise sources, $\alpha_{k}\;(\beta_{k})$
and $\alpha_{k}^{+}\;(\beta_{k}^{+})$ are not complex conjugate pairs,
except in the mean over a large number of stochastic integrations
of the above equations. These equations allow us to calculate
the expectation values of any desired time-normally ordered operator
moments as classical averages, exactly as required to calculate spectral correlations.

\section{Linearised analysis}
\label{sec:linearise}

In an operating regions where it is valid, a linearised fluctuation
analysis provides a simple way of calculating both intracavity and
output spectra of the system~\cite{DFW,mjc}, by treating it as an
Ornstein-Uhlenbeck process~\cite{ornstein}. To perform this analysis
we first divide the variables of \eref{eq:PPSDE} into a steady-state
mean value and a fluctuation part, e.g. $\alpha_{1}\rightarrow\alpha_{1}^{ss}+\delta\alpha_{1}$
and so on for the other variables. We find the steady state solutions
by solving the equations \eref{eq:PPSDE} without the noise terms (note that in this section we will treat all fields as being at resonance),
and write the equations for the fluctuation vector 
$\delta\tilde{x}=[\delta\alpha_{1},\delta\alpha_{1}^{+},\delta\alpha_{2},\delta\alpha_{2}^{+},\delta\beta_{1},\delta\beta_{1}^{+},\delta\beta_{2},\delta\beta_{2}^{+}]^{T}$,
to first order in these fluctuations, as 
\begin{equation}
\rmd\;\delta\tilde{x}=-A\delta\tilde{x}\; \rmd t +B\rmd W,
\label{eq:dAB}
\end{equation}
where the drift matrix is 
\begin{equation}
A=\left[\begin{array}{cc}
A_{aa} & -A_{ba}^{*}\\
A_{ba} & A_{bb}\end{array}\right],
\end{equation}
with
\begin{equation}
A_{aa} = \left[\begin{array}{cccc}
\gamma_{a}+i\Delta_{a} & -\kappa\beta_{1}^{ss} & -iJ_{a} & 0\\
-\kappa\beta_{1}^{ss\ast} & \gamma_{a}-i\Delta_{a} & 0 & iJ_{a}\\
-iJ_{a} & 0 & \gamma_{a}+i\Delta_{a} & -\kappa\beta_{2}^{ss}\\
0 & iJ_{a} & -\kappa\beta_{2}^{ss\ast} & \gamma_{a}-i\Delta_{a}
\end{array}\right],
\label{eq:Aaamat}
\end{equation}
and
\begin{equation}
A_{ba} = \left[\begin{array}{cccc}
\kappa\alpha_{1}^{ss} & 0 & 0 & 0\\
0 & \kappa\alpha_{1}^{ss\ast} & 0 & 0\\
0 & 0 & \kappa\alpha_{2}^{ss} & 0\\
0 & 0 & 0 & \kappa\alpha_{2}^{ss\ast}
\end{array}\right],
\label{eq:Abamat}
\end{equation}
and
\begin{equation}
A_{bb} = \left[\begin{array}{cccc}
\gamma_{b}+i\Delta_{b} & 0 & -iJ_{b} & 0\\
0 & \gamma_{b}-i\Delta_{b} & 0 & iJ_{b}\\
-iJ_{b} & 0 & \gamma_{b}+i\Delta_{b} & 0\\
0 & iJ_{b} & 0 & \gamma_{b}-i\Delta_{b}
\end{array}\right].
\label{eq:Abbmat}
\end{equation}

\begin{figure}
\begin{center}\includegraphics[width=0.4\columnwidth]{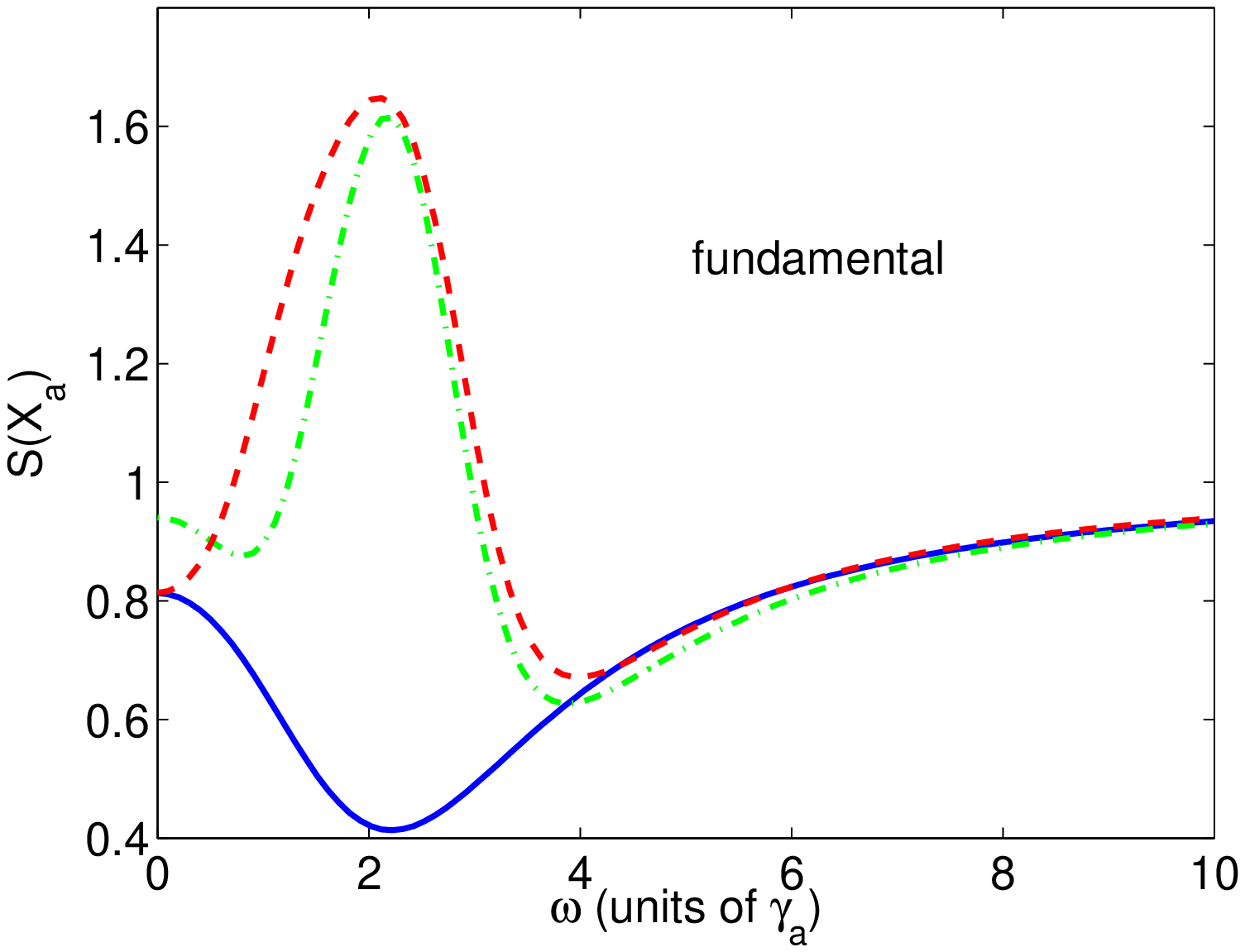}
\includegraphics[width=0.4\columnwidth]{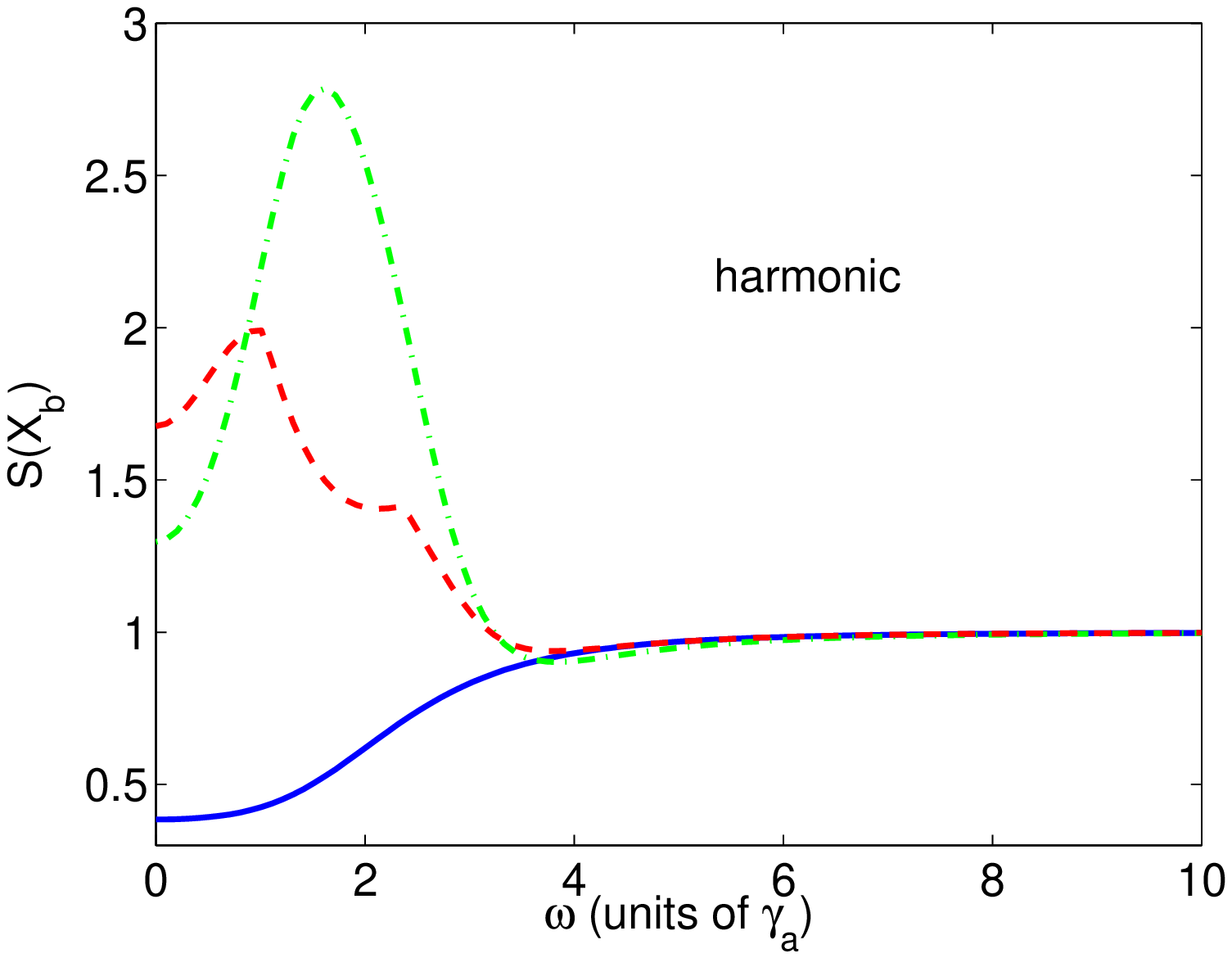}
\end{center}
\caption{The single-mode output variances for $\hat{X}_{a}$ and $\hat{X}_{b}$, for $\epsilon_{1}=\epsilon_{2}=0.8\epsilon_{c}$, $\Delta_{a}=\Delta_{b}=0$, $\gamma_{a}=\gamma_{b}$, and different values of $J_{a}$ and $J_{b}$. The solid lines are for the uncoupled case, with $J_{a}=J_{b}=0$. The dash-dotted lines are for $J_{a}=J_{b}=\gamma_{a}$, with the optimal quadrature angles being $\theta = 18^{o}$ (fundamental) and $13^{o}$ (harmonic). The dashed line is for $J_{a}=2\gamma_{a}=2J_{b}$ and $\theta=46^{o}$ (fundamental) and $74^{o}$ (harmonic). A value of less than one indicates squeezing. 
All values plotted here and in subsequent graphics are dimensionless.}
\label{fig:VXres}
\end{figure}

In \eref{eq:dAB}, $dW$ is a vector of real Wiener increments, and
the matrix $B$ is zero except for the first four diagonal elements,
which are respectively $\sqrt{\kappa\beta_{1}^{ss}},\;\sqrt{\kappa\beta_{1}^{ss\ast}},\;\sqrt{\kappa\beta_{2}^{ss}},\;\sqrt{\kappa\beta_{2}^{ss\ast}}$.
The essential conditions for this expansion to be valid are that moments
of the fluctuations be smaller than the equivalent moments of the
mean values, and that the fluctuations stay small. In the case of uncoupled SHG, it is well known that
there is a critical operating point above which this condition does
not hold and the system enters a self-pulsing regime~\cite{autopulse1,autopulse2,autopulse3}. This point is easily found by examination of the eigenvalues
of the equivalent fluctuation drift matrix for that system, and this
procedure is also valid in the present case. The fluctuations will
not tend to grow as long as none of the eigenvalues of the matrix
$A$ develop a negative real part. At the point at which this happens
the linearised fluctuation analysis is no longer valid, as the fluctuations
can then grow exponentially and the necessary conditions for linearisation
are no longer fulfilled. In this work we will only be interested in
a region where linearisation is valid and will restrict our analyses to below the self-pulsing threshold, found for \begin{equation}
\epsilon = \epsilon_{c} = \frac{2\gamma_{a}+\gamma_{b}}{\kappa}\sqrt{2\gamma_{b}(\gamma_{a}+\gamma_{b})}
\label{eq:critpump}
\end{equation}
in the case of a single uncoupled cavity.

\begin{figure}[tbhp]
\begin{center}\includegraphics[width=0.4\columnwidth]{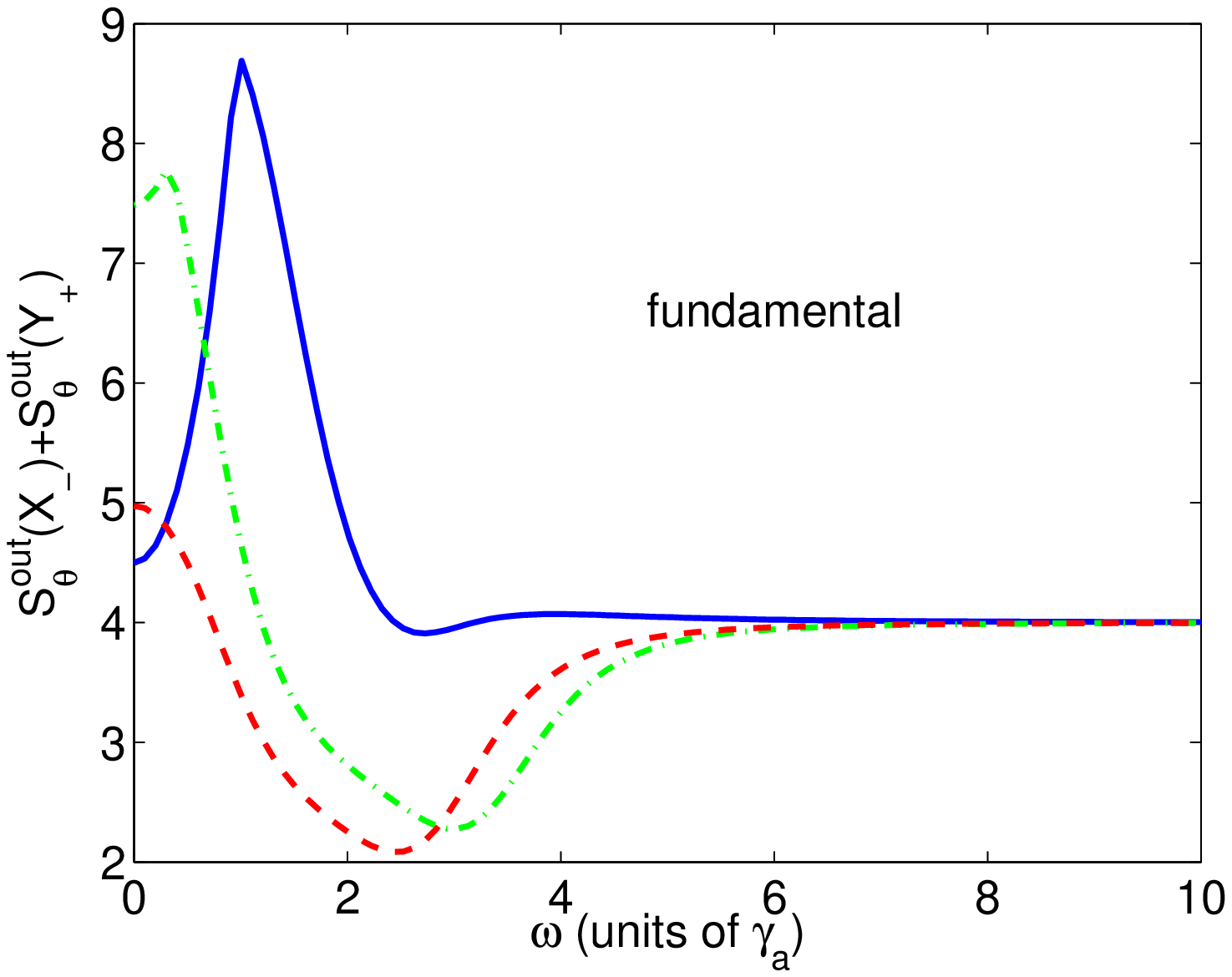}
\includegraphics[width=0.4\columnwidth]{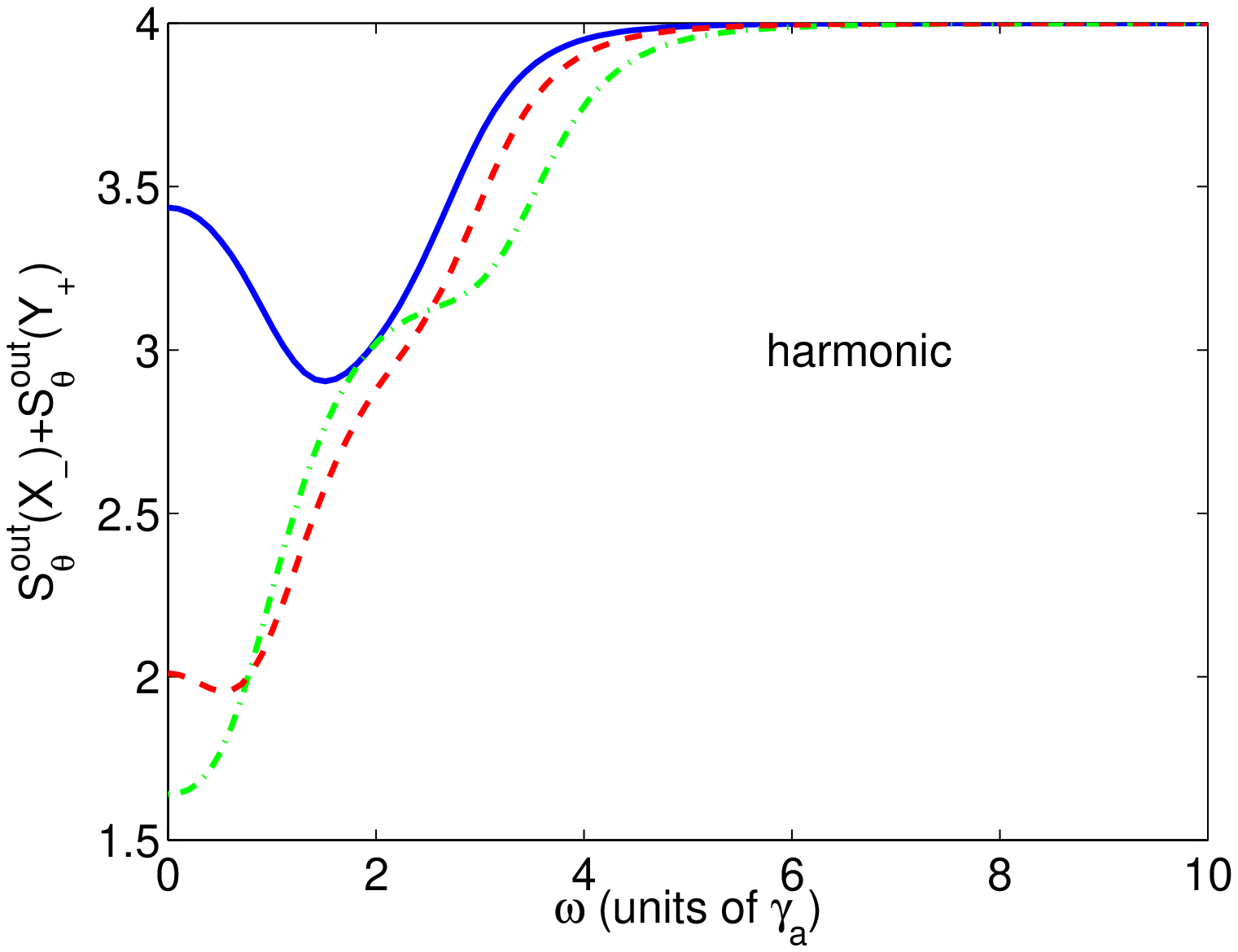}
\end{center}
\caption{The spectral correlations $S_{\theta}^{out}(\hat{X}_{-})+S_{\theta}^{out}(\hat{Y}_{+})$ of the fundamental (left) and harmonic (right), for $\epsilon_{1}=\epsilon_{2}=0.8\epsilon_{c}$, $\Delta_{a}=\Delta_{b}=0$ and different values of $J_{a}$ and $J_{b}$. The solid lines are for $J_{a}=J_{b}=\gamma_{a}$, with the optimal quadrature angles being $\theta = 63^{o}$ (fundamental) and $58^{o}$ (harmonic). The dash-dotted lines are for $J_{a}=2\gamma_{a}=2J_{b}$, at $\theta=91^{o}$ (fundamental) and $114^{o}$ (harmonic), while the dashed lines are for $J_{a}= 2\gamma_{a}$ and $J_{b}=\gamma_{a}/2$ at $\theta = 179^{o}$ (fundamental) and $119^{o}$ (harmonic). A value of less than $4$ indicates bipartite entanglement.}
\label{fig:Duanres}
\end{figure}

To examine the stability of the system, we first need to find the
steady state solutions for the amplitudes, by solving for the steady
state of \eref{eq:PPSDE} with the noise terms dropped. For this system in the general case, we find that it is convenient to solve for these steady states numerically, using a Runge-Kutta algorithm to integrate the system of deterministic equations until the solutions are well into the steady-state regime. The eigenvalues of the drift matrix are then also found numerically.
Using the steady-state solutions, we may then calculate any desired time normally-ordered
spectral correlations inside the cavity using the simple formula 
\begin{equation}
S(\omega)=\left(A+\rmi\omega 1\right)^{-1}BB^{T}\left(A^{T}-\rmi\omega 1\right)^{-1},
\label{eq:inspek}
\end{equation}
after which we use the standard input-output relations~\cite{mjc}
to relate these to quantities which may be measured outside the cavity. For example, the spectral output variance of the quadrature $\hat{X}_{1}^{\theta}\pm\hat{X}_{2}^{\theta}$ will be denoted $S_{\theta}^{out}(X_{\pm})$, with whether it refers to the fundamental or harmonic being made obvious by the context.  
In what follows we will use values of $\kappa=0.01$ and $\gamma_{a}=1$ while varying other parameters.

\begin{figure}[tbhp]
\begin{center}\includegraphics[width=0.4\columnwidth]{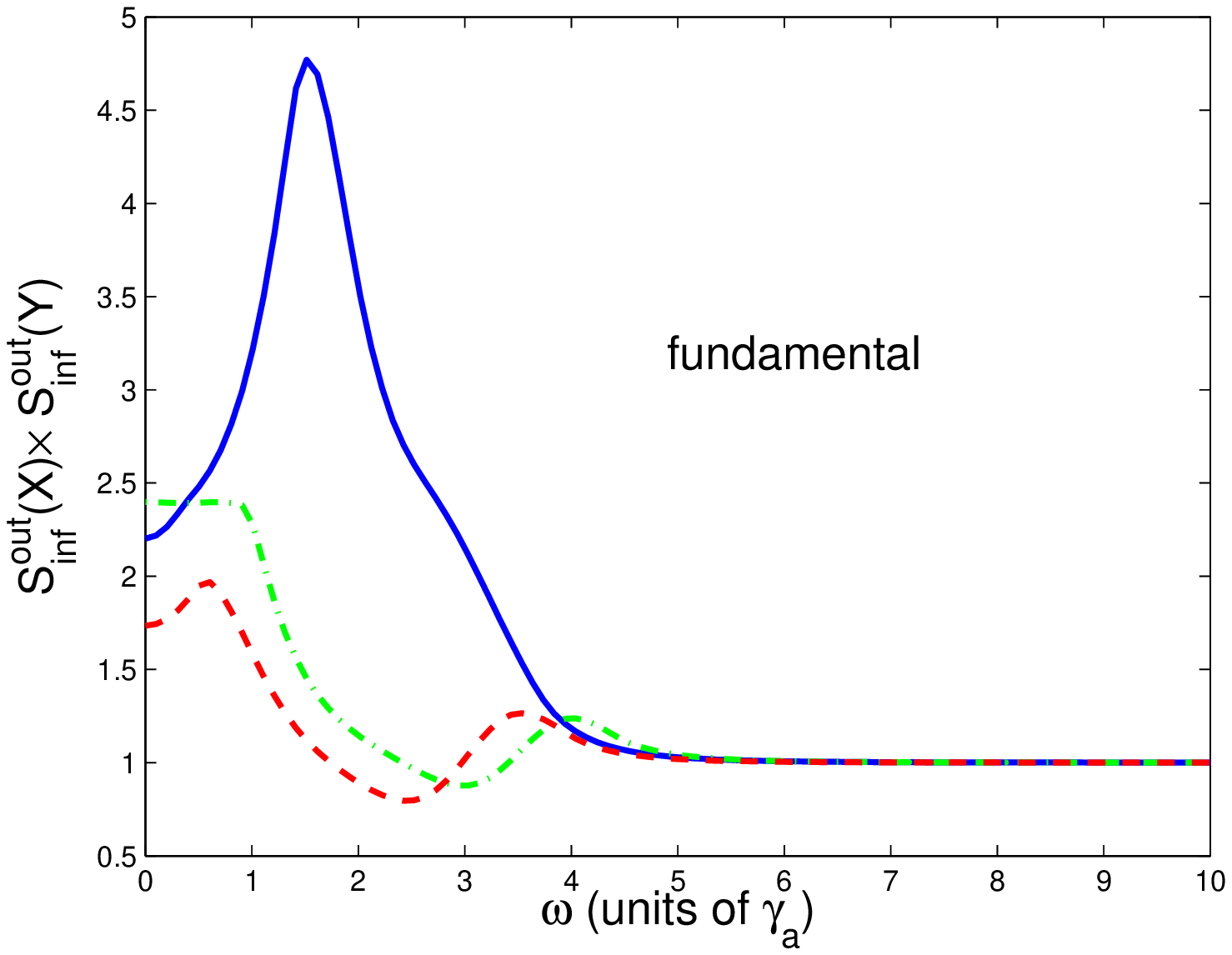}
\includegraphics[width=0.4\columnwidth]{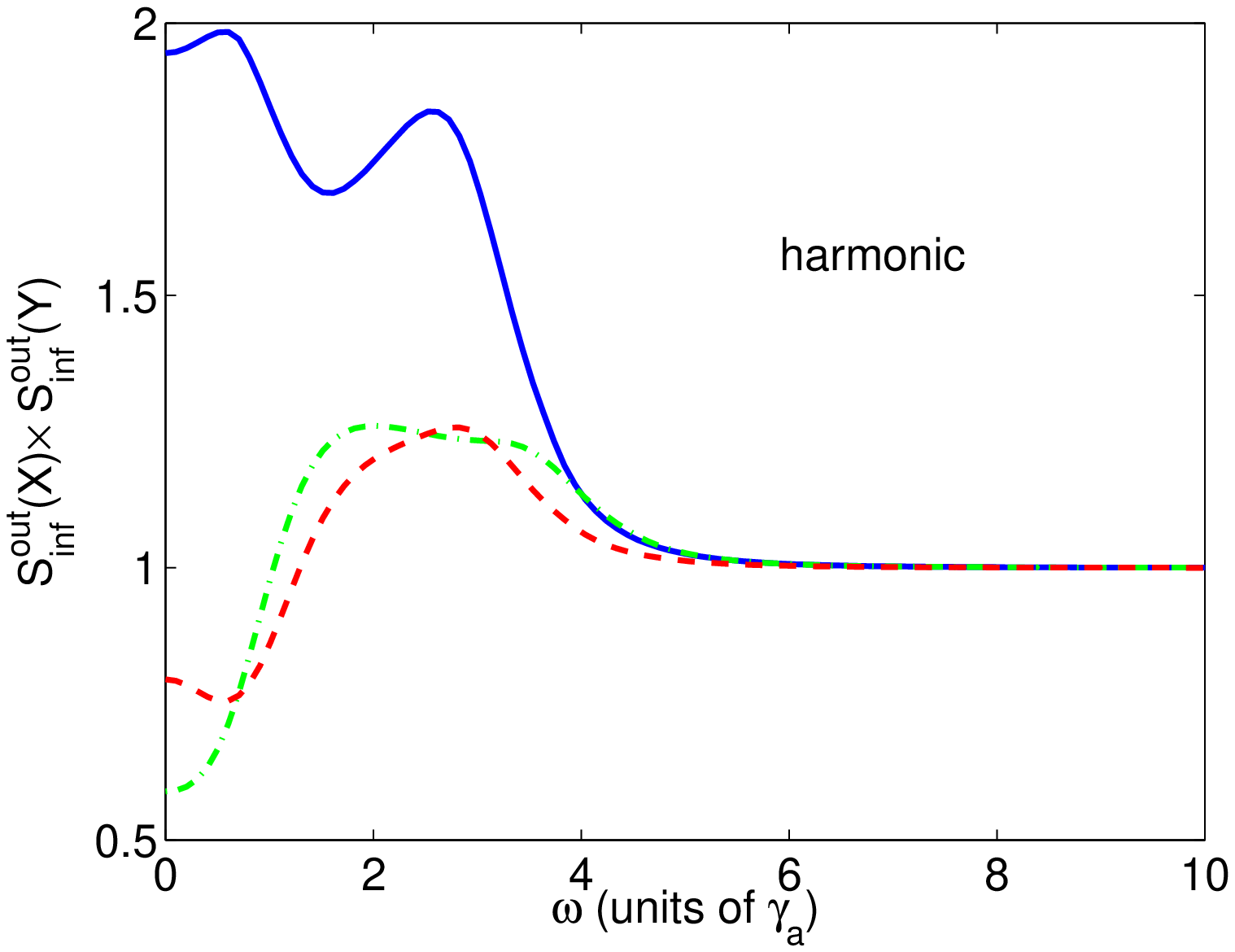}
\end{center}
\caption{The spectral correlations $S^{out}_{inf}(\hat{X})S^{out}_{inf}(\hat{Y})$ of both frequencies, for $\epsilon_{1}=\epsilon_{2}=0.8\epsilon_{c}$, $\Delta_{a}=\Delta_{b}=0$ and different values of $J_{a}$ and $J_{b}$. The solid lines are for $J_{a}=J_{b}=\gamma_{a}$, with the optimal quadrature angles being $\theta = 174^{o}$ (fundamental) and $45^{o}$ (harmonic). The dash-dotted lines are for $J_{a}=2\gamma_{a}=2J_{b}$ at $\theta=46^{o}$ (fundamental) and $24^{o}$ (harmonic), while the dashed lines are for $J_{a}= 2\gamma_{a}$ and $J_{b}=\gamma_{a}/2$ at $\theta = 44^{o}$ (fundamental) and $29^{o}$ (harmonic). A value of less than $1$ illustrates the EPR paradox and hence bipartite entanglement. We see that this measure is markedly less sensitive to entanglement at the lower frequency than the correlations of \fref{fig:Duanres}.}
\label{fig:EPRres}
\end{figure}

\section{Doubly resonant cavity}
\label{sec:duplares}

In this section we will give results for the case where all four intracavity fields are at resonance. The first correlations we show, in \fref{fig:VXres} are the minimum quadrature variances for each mode for different values of the coupling strengths. We see that the quadrature angle of minimum noise is changed by comparison with the resonant case, and that the harmonic exhibits no single-mode squeezing for the coupling strengths shown. The single-mode squeezing in the fundamental is also noticeably degraded from the uncoupled configuration, and both show excess noise at some frequencies. This is common with entangled systems, where the modes considered individually will show excess noise and the entanglement manifests itself in the correlations of this noise between the two modes. The fact that the minimum values of the correlations are no longer found at either $\theta=0$ or $\pi/2$ is a consequence of the evanescent coupling, and has previously been seen in the coupled downconversion configuration~\cite{mkopdd,Nicolas}. 

\begin{figure}[tbhp]
\begin{center}
\includegraphics[width=0.4\columnwidth]{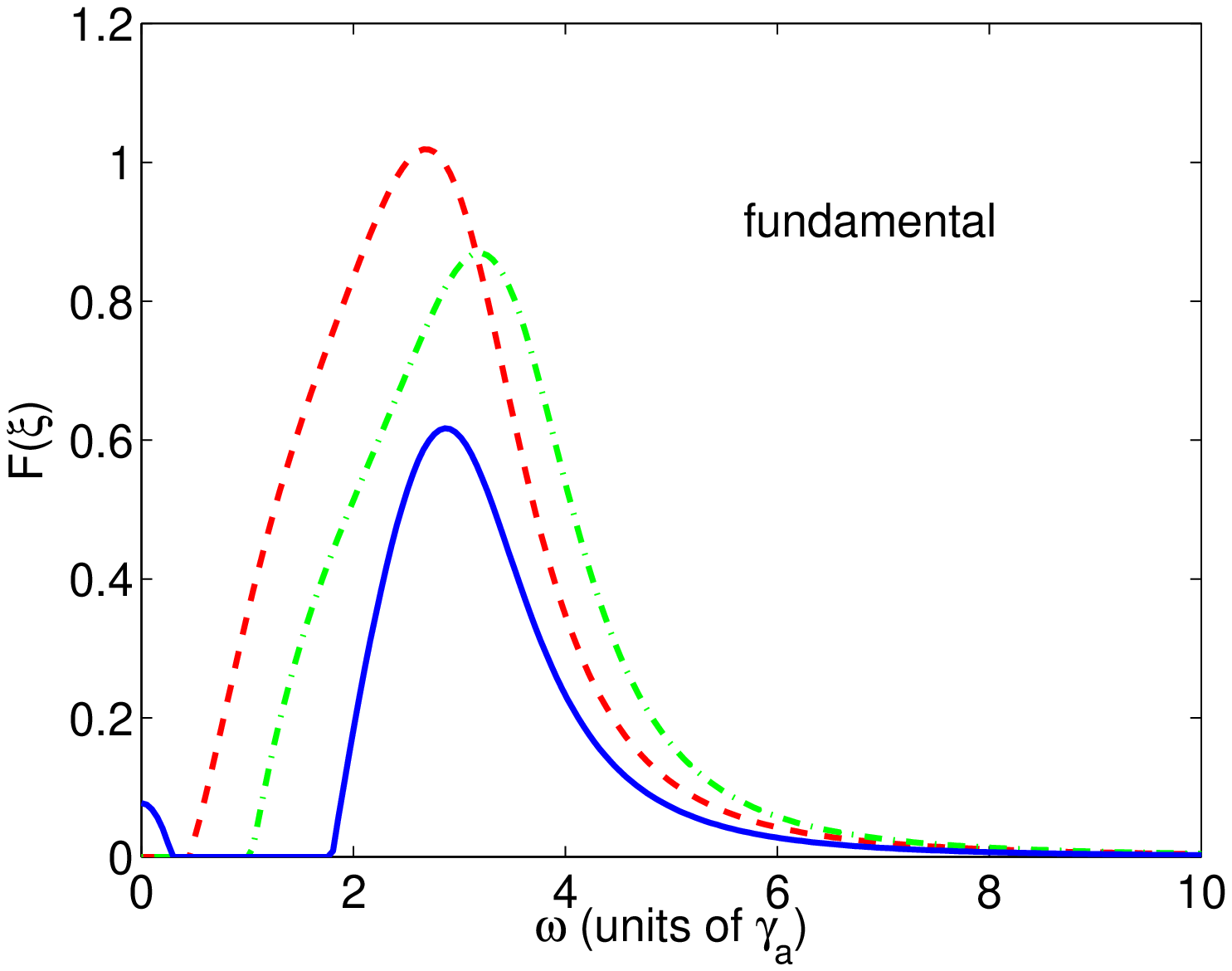}
\includegraphics[width=0.4\columnwidth]{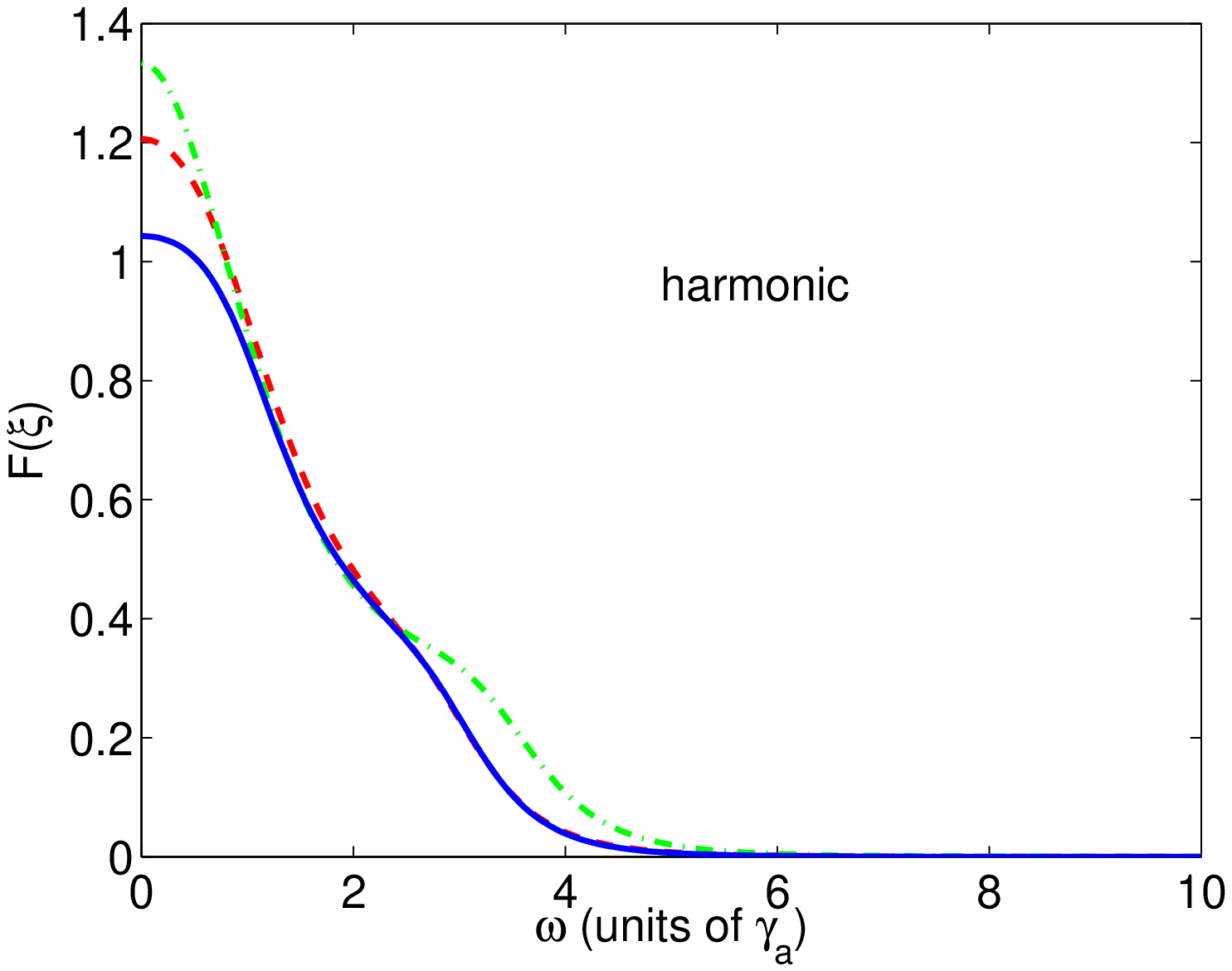}
\end{center}
\caption{The logarithmic negativity, $\cal{F}(\xi)$, for $\epsilon_{1}=\epsilon_{2}=0.8\epsilon_{c}$, $\Delta_{a}=\Delta_{b}=0$ and different values of $J_{a}$ and $J_{b}$. The left plot shows the values for the low frequency modes and the right plot shows them for the high frequency modes. 
The solid lines are for $J_{a}=J_{b}=\gamma_{a}$, the dash-dotted lines are for $J_{a}=2\gamma_{a}=2J_{b}$, and the dashed lines are for $J_{a}= 2\gamma_{a}$ and $J_{b}=\gamma_{a}/2$.}
\label{fig:lognegres}
\end{figure}

In the next three figures we examine the Duan and Simon criteria, the EPR criteria and the logarithmic negativity as the couplings are changed. In \fref{fig:Duanres} we show the minima of the correlations at the quadrature angles for which these are found. We see that the angles are quite different for the fundamental and harmonic for the same parameters and that the degree of violation of the inequalities varies markedly with coupling strengths. The fact that the quadrature angles change should not be a problem experimentally, as homodyne detection techniques generally scan through all angles. When we examine \fref{fig:EPRres}, we see that the EPR criteria do not provide a very sensitive measure of the presence of entanglement for the fundamental modes but are better for the harmonic, although the entanglement between the two harmonic modes for $J_{a}=J_{b}=\gamma_{a}$ is completely missed. This is not in contradiction with the results shown in \fref{fig:Duanres} because the EPR criteria are merely sufficient but not necessary to demonstrate inseparability of the modes. It is also apparent that they give their best results at different quadrature angles to the Duan and Simon criteria.

The logarithmic negativity is perhaps the most sensitive indicator of inseparability for this system, although it does not tell us the quadrature angle at which maximum violations of the other inequalities can be found. In \fref{fig:lognegres}, we see, for example, that it is much more sensitive at finding entanglement between the fundamental modes than the other two methods in the case where $J_{a}=J_{b}=\gamma_{a}$. We note here that the Duan and Simon criteria showed very little violation of the inequality in this case, but we have shown only the quadrature angle of maximum violation. The frequency range over which the logarithmic negativity is positive tells us that violations can be found for other quadrature angles as the frequency changes, although given the fact that the maximum violation of the Duan and Simon inequality is very small here, it is unlikely that this particular combination of parameters would be a good operational choice.

\section{Detuning the cavity}
\label{sec:detune}

Often in optical systems the best performance is found when the cavity
is resonant for the different modes involved in the interactions, with any detuning worsening the quantum features such as squeezing and changing the quadrature angles at which these are seen~\cite{granja}.
In the present case we find that detuning the cavity by the appropriate
amount from the two frequencies allows for some simplification of
the theoretical analysis and can actually improve some quantum correlations. In the special case where $\Delta_{a}=J_{a}$ and $\Delta_{b}=J_{b}$ the best quadrature angles remain fixed as the pumping is varied. 
If we also set $\epsilon_{1}=\epsilon_{2}=\epsilon$, the equations become somewhat simplified and we are able to progress further analytically. In this case we define the new steady-state variables, $\alpha_{\pm}=\alpha_{1}\pm\alpha_{2}$ and $\beta_{\pm}=\beta_{1}\pm\beta_{2}$ and solve the classical equations for these. Due to the symmetry of the system, we may assume that $\alpha_{-}=\beta_{-}=0$, and that the variables are real. Numerical analysis bears out these assumptions. We then find that
\begin{equation}
\beta_{+} = -\frac{\kappa}{4\gamma_{b}}\alpha_{+}^{2},
\label{eq:betaplus}
\end{equation}
with $\alpha_{+}$ being the real solution of the cubic equation
\begin{equation}
\frac{\kappa^{2}}{8\gamma_{b}}\alpha_{+}^{3}+\gamma_{a}\alpha_{+}-2\epsilon = 0.
\label{eq:alphaplus}
\end{equation}
Setting
\begin{equation}
\chi = \left(9\kappa^{4}\gamma_{b}\epsilon+\sqrt{24\kappa^{6}\gamma_{a}^{3}\gamma_{b}^{3}+
81\kappa^{8}\gamma_{b}^{2}\epsilon^{2}}\right)^{1/3},
\label{eq:chisub}
\end{equation}
we find
\begin{equation}
\alpha_{+} = \frac{2\chi}{3^{2/3}\kappa^{2}}-\frac{4\gamma_{a}\gamma_{b}}{3^{1/3}\chi}.
\label{eq:solaplus}
\end{equation}
We now need to solve for the fluctuations in the new variables. Setting $\alpha_{1}+\alpha_{2}=\alpha_{+}+\delta\alpha_{+}$ etc, we write the equations of motion for the variables $\delta\tilde{x}_{\pm}=[\delta\alpha_{+},\delta\alpha_{+}^{+},\delta\alpha_{-},\delta\alpha_{-}^{+},\delta\beta_{+},\delta\beta_{+}^{+},
\delta\beta_{-},\delta\beta_{-}^{+}]^{T}$ in matrix form as 
\begin{equation}
\delta\tilde{x}_{\pm} = -A_{pm}\delta\tilde{x}_{\pm}dt+B_{\pm}dW,
\label{eq:plusminusmat}
\end{equation}
where
\begin{equation}\fl
A_{pm}=\left[
\begin{array}{cccccccc}
\gamma_{a} & -\frac{\kappa}{2}\beta_{+} & 0 & 0 & -\frac{\kappa}{2}\alpha_{+}^{\ast} & 0 & 0 & 0 \\
-\frac{\kappa}{2}\beta_{+}^{\ast} & \gamma_{a} & 0 & 0 & 0 & -\frac{\kappa}{2}\alpha_{+} & 0 & 0 \\
0 & 0 & \gamma_{a}+2\rmi J_{a} & 0 & 0 & 0 & 0 & 0 \\
0 & 0 & 0 & \gamma_{a}-2\rmi J_{a} & 0 & 0 & 0 & 0 \\
\frac{\kappa}{2}\alpha_{+} & 0 & 0 & 0 & \gamma_{b} & 0 & 0 & 0 \\
0 & \frac{\kappa}{2}\alpha_{+}^{\ast} & 0 & 0 & 0 & \gamma_{b} & 0 & 0 \\
0 & 0 & 0 & 0 & 0 & 0 & \gamma_{b}+2\rmi J_{b} & 0 \\
0 & 0 & 0 & 0 & 0 & 0 & 0 & \gamma_{b}-2\rmi J_{b}
\end{array}\right],
\label{eq:Aplusminus}
\end{equation}
and
\begin{equation}
B_{pm} = \left[
\begin{array}{cc}
B_{4} & B_{0} \\
B_{0} & B_{0}
\end{array}\right],
\end{equation}
where 
\begin{equation}
B_{4}=\left[
\begin{array}{cccc}
\sqrt{\frac{\kappa}{2}\beta_{+}} & 0 & \sqrt{\frac{\kappa}{2}\beta_{+}} & 0\\
0 & \sqrt{\frac{\kappa}{2}\beta_{+}^{\ast}} & 0 & \sqrt{\frac{\kappa}{2}\beta_{+}^{\ast}}\\
\sqrt{\frac{\kappa}{2}\beta_{+}} & 0 & -\sqrt{\frac{\kappa}{2}\beta_{+}} & 0\\
0 & \sqrt{\frac{\kappa}{2}\beta_{+}^{\ast}} & 0 & -\sqrt{\frac{\kappa}{2}\beta_{+}^{\ast}}
\end{array}\right],
\label{eq:Bplusminus}
\end{equation}
and the $B_{0}$ are $4\times 4$ null matrices.

%\subsection{Stability and spectra}

In this case we can find analytical solutions for the eigenvalues of $A_{pm}$, which we write in terms of $\alpha_{+}$ and $\beta_{+}$ as
\begin{eqnarray}\fl
\eqalign{
\lambda_{1,2} &= \frac{1}{8}\left[4\Gamma-2\kappa\beta_{+}\pm\sqrt{\left[4\Gamma-2\kappa\beta_{+}\right]^{2}-16\left(4\gamma_{a}\gamma_{b}-2\gamma_{b}\kappa\beta_{+}+\kappa^{2}\alpha_{+}^{2}\right)}\right],\\
\lambda_{3,4} &= \frac{1}{8}\left[4\Gamma+2\kappa\beta_{+}\pm 2\sqrt{\left[2\Gamma+\kappa\beta_{+}\right]^{2}-4
\left[4\gamma_{a}\gamma_{b}+2\gamma_{b}\kappa\beta_{+}+\kappa^{2}\alpha_{+}^{2}\right]}\right],\\
\lambda_{5,6} &= \gamma_{a}\pm 2\rmi J_{a}, \\
\lambda_{7,8} &= \gamma_{b}\pm 2\rmi J_{b},
}
\label{eq:pmflucs}
\end{eqnarray}
where $\Gamma = \gamma_{a}+\gamma_{b}$.

\begin{figure}[tbhp]
\begin{center}\includegraphics[width=0.75\columnwidth]{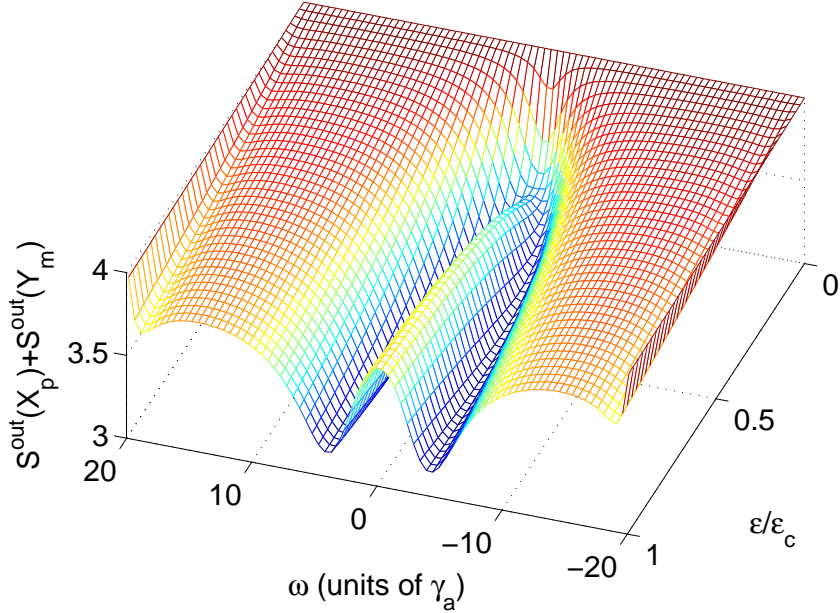}
\end{center}
\caption{The output spectral correlation $S^{out}(X_{p})+S^{out}(Y_{m})$ of the low frequency modes, for $\kappa=0.01$, $\gamma_{a}=1$, $\gamma_{b}=2$, $\Delta_{a}=J_{a}=10\gamma_{a}$, and $\Delta_{b}=J_{b}=2\gamma_{a}$.}
\label{fig:pmDuanA}
\end{figure}

There are two types of instability which we may expect in this system and which would invalidate a linearised fluctuation analysis. The first is when one or more of the eigenvalues above has a negative real part and the second is the self-pulsing regime~\cite{autopulse1,autopulse2,autopulse3}, which begins where there exist complex conjugate eigenvalues with real part equal to zero. Examining the expressions of \eref{eq:pmflucs}, we see that the last four can never cause any problems, but that some of the others could develop negative real parts. While analytical solutions for these in terms of the pump strength rather than the cavity field values can be found, these are extremely unwieldy. However, numerical analysis shows that the system is stable up to the same critical pumping as given in \eref{eq:critpump}, so we will give results in this regime. 

\begin{figure}[tbhp]
\begin{center}\includegraphics[width=0.75\columnwidth]{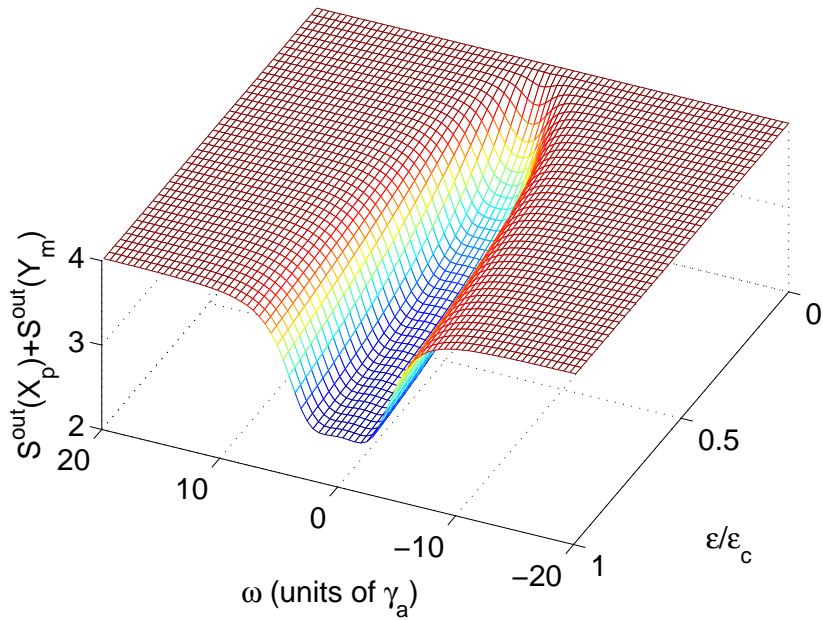}
\end{center}
\caption{The output spectral correlation $S^{out}(X_{p})+S^{out}(Y_{m})$ of the high frequency modes, for $\kappa=0.01$, $\gamma_{a}=1$, $\gamma_{b}=2$, $\Delta_{a}=J_{a}=10\gamma_{a}$, and $\Delta_{b}=J_{b}=2\gamma_{a}$.}
\label{fig:pmDuanB}
\end{figure}

In terms of the quadrature variances used in section~\ref{sec:linearise},
we now define 
\begin{eqnarray}
\eqalign{
X_{p} &= A_{p}+A_{p}^{+}=X_{1}+X_{2}, \\
X_{m} &= A_{m}+A_{m}^{+}=X_{1}-X_{2},\\
Y_{p} &= -i\left(A_{p}-A_{p}^{+}\right)=Y_{1}+Y_{2}, \\
Y_{m} &= -i\left(A_{m}-A_{m}^{+}\right)=Y_{1}-Y_{2},}
\label{eq:quadcombine}
\end{eqnarray}
and similarly for the second harmonic quadratures, so that
we can give expressions for the output spectral variances of these new
quadratures. Because of the way the variables are defined, the easiest correlations to extract are those of \eref{eq:critDuan}, which can be constructed from the variances of the four quadratures defined above, with
\begin{eqnarray}
\eqalign{
V(X_{1}-X_{2})+V(Y_{1}+Y_{2}) &= V(X_{m})+V(Y_{p}), \\
V(X_{1}+X_{2})+V(Y_{1}-Y_{2}) &= V(X_{p})+V(Y_{m}).
}
\label{eq:newDuan}
\end{eqnarray}
We find that, at least in the parameter regimes we have investigated numerically, it is the second of these correlations which violate the inequalities, which is different to the situation with, for example, coupled downconverters~\cite{mkopdd}.

\begin{figure}[tbhp]
\begin{center}\includegraphics[width=0.75\columnwidth]{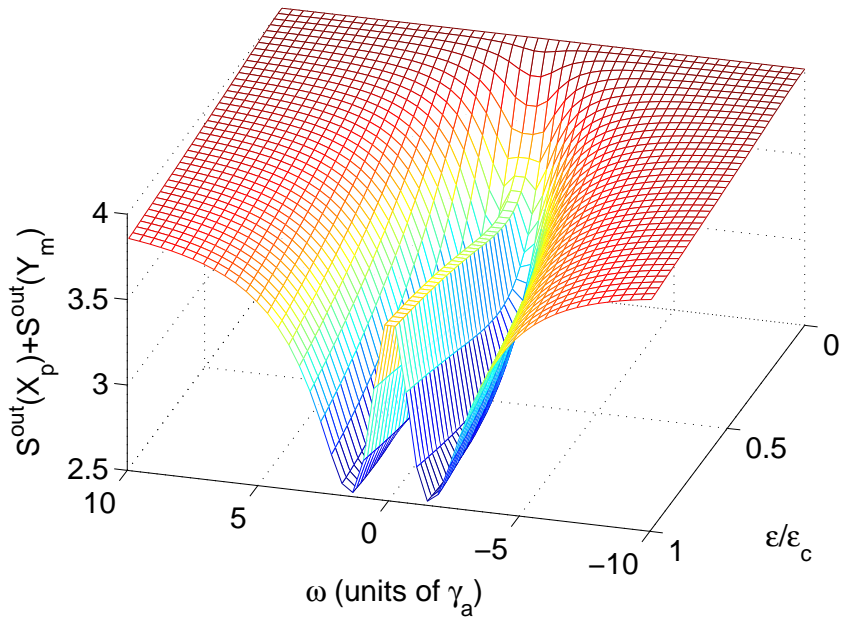}
\end{center}
\caption{The output spectral correlation $S^{out}(X_{p})+S^{out}(Y_{m})$ of the low frequency modes, for $\kappa=0.01$, $\gamma_{a}=1$, $\gamma_{b}=\gamma_{a}/2$, $\Delta_{a}=J_{a}=10\gamma_{a}$, and $\Delta_{b}=J_{b}=2\gamma_{a}$.}
\label{fig:pmDuanA10}
\end{figure} 

\begin{figure}[tbhp]
\begin{center}\includegraphics[width=0.75\columnwidth]{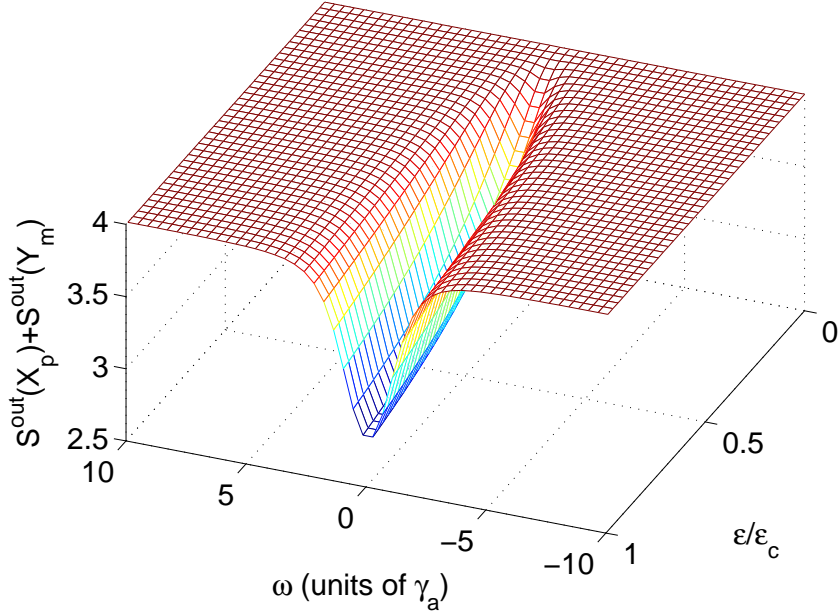}
\end{center}
\caption{The output spectral correlation $S^{out}(X_{p})+S^{out}(Y_{m})$ of the high frequency modes, for $\kappa=0.01$, $\gamma_{a}=1$, $\gamma_{b}=\gamma_{a}/2$, $\Delta_{a}=J_{a}=10\gamma_{a}$, and $\Delta_{b}=J_{b}=2\gamma_{a}$.}
\label{fig:pmDuanB10}
\end{figure}

\begin{figure}[tbhp]
\begin{center}\includegraphics[width=0.75\columnwidth]{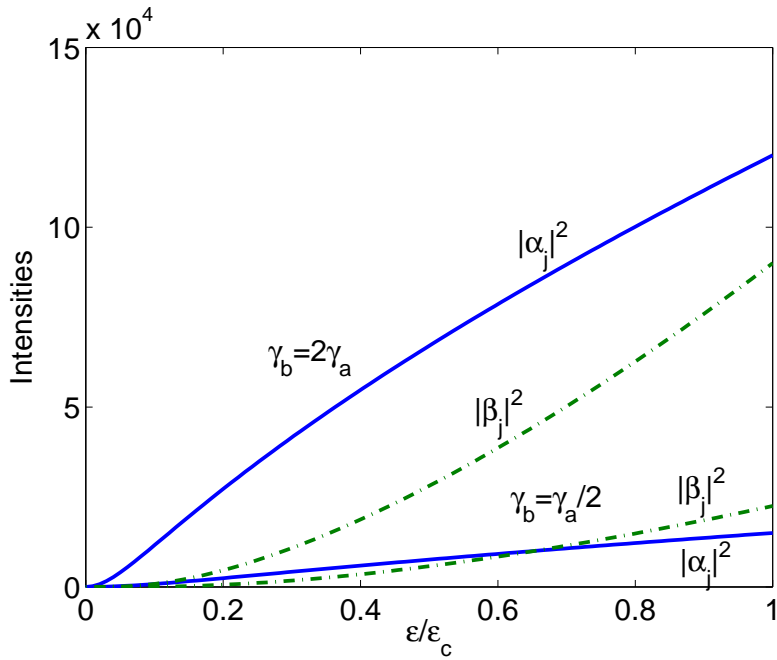}
\end{center}
\caption{The intracavity intensities for each mode as a function of the ratio $\epsilon/\epsilon_{c}$. The upper pair and the lower pair are for different ratios of the cavity loss rates at each frequency, with $\kappa=0.01$, $\gamma_{a}=1$, $\Delta_{a}=J_{a}=10\gamma_{a}$, and $\Delta_{b}=J_{b}=2\gamma_{a}$.}
\label{fig:intensidade}
\end{figure}

We show spectral results obtained numerically as the pumping strength is varied up to its critical value, and for different values of the cavity loss rates. In \fref{fig:pmDuanA} and \fref{fig:pmDuanB} we present results for the loss rate at the harmonic frequency being twice that of the fundamental, showing that the violations increase as the pumping increases up to the critical value. In \fref{fig:pmDuanA} we see that the spectrum bifurcates and note that this could be advantageous experimentally as the region around zero frequency is often swamped by technical noise. In \fref{fig:pmDuanA10} and \fref{fig:pmDuanB10} we give the spectra for the case where the loss rate at the fundamental frequency is twice that of the harmonic. The main difference is that the violations of the inequalities happen over a narrower range of frequencies, with both violations again increasing as the pump increases. In comparison with coupled downconversion below threshold~\cite{mkopdd}, the entanglement produced here is with reasonably intense fields, as shown in \fref{fig:intensidade}, which could be a real advantage in some applications. 

\section{Conclusions}

We have analysed the system of intracavity evanescently coupled second harmonic generation in terms of phase sensitive correlations which give evidence of continuous variable entanglement between different modes of the optical field. We have shown how matching the cavity detunings to the evanescent coupling rates fixes the quadrature angles for which the best violations of the inequalities occur. 
We see that below the self-pulsing threshold, the system exhibits a wide range of behaviour depending on the relative strengths of the pumps, the cavity loss rates, the detunings and the evanescent couplings. As all of these are experimentally tuneable, the device may be of use for applications which require varying degrees of entanglement to be available at different intensities, different frequencies and different quadrature phase angles. The entangled beams exit the cavity at different spatial locations and do not have to be separated before measurements can be made. As the system also produces single mode squeezing and can be built using integrated optics, it may prove to be more robust and useful than devices which rely on the relative stability and positioning of individual optical components. Finally, the fact that the entanglement is present with reasonably intense fields may prove to be a real advantage over devices based on nondegenerate downconversion, which experience phase diffusion in the region where the fields develop macroscopic intensities.

\ack

This research was supported by the Australian Research Council and the Queensland state government.

\end{document}